# Position Paper: Provenance Data Visualisation for Neuroimaging Analysis

Bilal Arshad, Kamran Munir, Richard McClatchey & Saad Liaquat
*Centre for Complex Cooperative Systems, University of the West of England, Bristol, UK.*
*{Bilal.Arshad, Kamran2.Munir, Richard.McClatchey, Saad2.Liaquat} @uwe.ac.uk*

*Abstract*- Visualisation facilitates the understanding of scientific data both through exploration and explanation of visualised data. Provenance contributes to the understanding of data by containing the contributing factors behind a result. With the significant increase in data volumes and algorithm complexity, clinical researchers are struggling with information tracking, analysis reproducibility and the verification of scientific output. Data coming from various heterogeneous sources (multiple sources with varying level of trust) in a collaborative environment adds to the uncertainty of the scientific output. Systems are required that offer provenance data capture and visualisation support for analyses. We present an account for the need to visualise provenance information in order to aid the process of verification of scientific outputs, comparison of analyses, progression and evolution of results for neuroimaging analysis.

## I. Introduction

In a large-scale distributed research environment it is essential to develop systems that support and manage provenance data. This alone may not be enough for researchers to fully understand and accept the results, being missing are data sources and data processing services used to derive the scientific data and intermediate data produced during the derivation process. In other words a researcher cannot know the reason behind a scientific result unless they see a picture of the complete and complex cause and effect. In fact a researcher may need to visually see the combination of scientific data and its provenance, which together gives the meta-data to ascertain the results. Scientific workflows are increasingly becoming popular or orchestrate research processes in medical analyses, to ensure the reproducibility of results and to confirm the correctness of the outcomes [1]. In a collaborative research environment, where researchers use each other's results and methods, traceability of data generated, stored and used must also be maintained. All these forms of knowledge are collectively referred to as forms of so-called 'provenance' information.

The availability of provenance information is as important as the results of the scientific analysis itself [2]. In any system where there are multiplicities of datasets, and version of workflows operating upon those data set, particularly when the analysis is carried out repetitively and/or in collaborative teams, it is imperative to retain a record of who did what, to which dataset, on which dates, as well as recording the outcome(s) of the analysis. This 'provenance' information needs to be logged as records of 'particular users' analyses so that they can be reproduced and amended and repeated as part of a robust research process. All of this information, normally generated through execution of scientific workflows enable the traceability of the origins of data (and processes); can identify event causality; enable broader forms of sharing, reuse, and long-term preservation of scientific data; can be used to attribute ownership and determine the quality of particular data set [3].

Provenance simply means the history, ownership and usage of data and its processing in some domain of interest. For example the logging of process execution in the study of High Energy Physics (HEP) experiments at CERN. In order to verify and interpret the results produced by scientific analysis of this data, researchers require reliable provenance information [4]. Similarly, in order to assist research into diseases such as Alzheimer's disease, researchers require scientific workflows (a.k.a pipelines) to process brain scans for various biomarkers. These biomarkers include the cortical thickness of a brain, the thinning of which has been linked with the onset of Alzheimer's disease. Researchers can track the progression of the disease by employing image analysis algorithms into neuroimaging workflows. The knowledge acquired from executing these neuroimaging workflows must be validated using provenance information. In the health informatics community great emphasis has been placed on the provision of infrastructures to support biomedical researchers for the purpose of data capture, image analysis, and the processing of scientific workflows and the sharing of diagnosis. This may include browsing data samples and specifying and executing workflows (or pipelines) of algorithms required for neurological analysis. Visualisation and provenance techniques, although used rarely in combination, may further help to increase the scientist's ability to understand scientific results since the scientist may be able to use a single tool to evaluate final results, the derivation process and any intermediate results produced during the experiment. In order to aid the researchers in the exploration process there is a need to visualise the data product and the associated provenance data.

In the above context, this paper aims to highlight the need for provenance visualisation for neuroimaging analysis. This will provide the emphasis on the need for clinical researchers to generate visualisations to aid the exploration process by providing them with complete visualisation of the data product and the associated provenance. Existing state of the art workflow systems are not completely generic and reconfigurable. Most workflow provenance management systems are designed for data-flow oriented workflows and researchers are now realising that tracking data alone is insufficient to support the scientific process (for example, see [5]). As a starting point, we take our user requirements based on NeuGRID Project, which started in January 2010 to provide computing and storage infrastructure, and services to store neuroimages and to facilitate neuro-researchers in defining and executing neuro analysis on stored neuroimages. The remainder of this paper is structured as follows: Section 2, need for provenance visualisation, with emphasis on neuroimaging analysis; Section 3, requirements for provenance visualisation; Section 4, NeuroProv System Architecture; Section 5, Use Cases; Section 6, Research Methodology and Section 7, Conclusions and Future Research.

## II. Provenance Visualisation

The practice of representing information visually is not new. Students to scientists, analysts to politicians, all over the world have been using data visualisation to track everything from DNA sampling to stock prices. According to Friedman [6], data visualisation's main goal is to communicate information clearly and effectively through graphical means. It does not mean that data visualisation needs to look uninteresting to be functional or extremely sophisticated to look beautiful. To convey ideas effectively, both aesthetic form and functionality need to go hand in hand, providing insights into a rather sparse and complex data set by communicating its key aspects in a more intuitive way.

Scientists rely on visualisations to aid in data exploration; which is often a complex process that requires close collaboration among domain scientists, computer scientists and visualisation experts. The ability to collaboratively explore data is one key to the scientific discovery process. The domain of neuroimaging analysis is no such exception. Neuroimaging is a crucial tool for both research and clinical neuroscience. In order to assist research into various neurodegenerative diseases such as Alzheimer's, researchers require to process brain scans for various biomarkers. These biomarkers include the cortical thickness of the brain, thinning of which has been linked to the onset of Alzheimer's disease. A significant challenge in neuroimaging and in fact all biological sciences, concerns devising ways to manage the enormous amounts of data generated using current techniques. This challenge is compounded by the expansion of collaborative efforts in recent years and the necessity of not only sharing data across multiple sites, but making that data available and useful to the scientific community at large.

Scientific experiments such as DNA Analysis [7], those at the Large Hadron Collider (LHC) [8] and projects such as NeuGRID [9] generate extremely large amounts of data. These communities use scientific workflows [10] to orchestrate the complex processing of data for their analyses. During the computation of this large pool of data, scientists end up creating an even larger pool of data representing intermediate results and associated metadata. An important consideration during data processing is that of understanding these intermediate results and the processes involved in order to derive the final result, to verify the authenticity of the results produced and to provide insight.

Prior visualisation systems tend to deal with either the data product or the process but not both. Specifically, Taverna [11] uses visualisation to help answer questions that establish how the experiment results were obtained; VisTrails [12] allows user to navigate workflow version in an intuitive way, to visually compare different workflow and their results, and to examine the actions that led to the result; Probe-It! [13] enable scientists to move the visualisation focus from intermediate and final results to provenance back and forth; the Prototype Lineage Server [14] allows users to browse lineage information by navigating through sets of metadata that provide useful details about the data products and transformations in a workflow invocation; Pedigree Graph [15], one of the tools in Multi-Scale Chemistry (MSC) portal from the Collaboratory of Multi Scale Chemical Science (CMCS), is designed to enable users to view multi-scale data provenance; the MyGrid project renders graph-based views of RDF-coded provenance using Haystack [16].

## III. User Requirements for Provenance Visualisation

Kunde *et. al.* [17] derive abstract user requirements for provenance visualisation, including: 1)process: the sequence of process steps is the centre of inspection; 2)results: the intermediate or end results of interactions are the centre of users view; 3)relationship: the relationship between actors is important; 4)timeline: the time is important to observe; 5)participation: the correctness of the participants is important; 6)compare: the comparison of subjects shows the difference between then; 7) interpretation: an individual visualisation view depending upon end-user's requirement.

The goal of our visualisation research is to serve both the broad and narrowly focused audiences in the domain of neuroimaging analysis, so it addresses each of the above requirement as follows: 1-3) our visualisation tool is based upon an accepted model for provenance representation, namely, the Open Provenance Model (OPM) [18], which denotes entities (processes, artifacts and agents) as nodes, and the relationship between them as edges in a graph. It is able to show complete graph with both the process steps and intermediate (final) results, or abstract graphs focusing on either one of them; 4) the OPM is capable of representing time information to nodes and edges; 5) participation is represented by agents through "wasControlledBy" relationship in the OPM, so our tool helps the user visually evaluate the correctness of participation; 6) users can compare attributes of nodes using our tool, the users can also use it to compare two graphs; 7) for the last type of user requirement (interpretation), we aim to show how we satisfy it with a customised view of the graph based on our users access privileges/requirements. (For a more detailed summary on OPM, we refer our reader to [18]).

Goble et al., [19] define the seven W's (Who, What, Where, Why, When, Which, (W) how) in order to encompass aspects of provenance. This ascertain features of provenance fundamental for the use of provenance data such as the person involved in the experiment (who); the material and methods used in the experiment (what and how); the conditions and timings at the time of the experiment (when and where); the purpose of running the experiment (why); as well as the results and the conclusions of the experiment (what). For researchers and scientists in the domain of neuro-imaging the intention and the results of experiments are of crucial importance but also the understanding of "how to" of experiments. Based on Goble's seven W's, generic requirements defined in [7] and using the N4U as a case study we assert the following. Scientists can use the visualised provenance for the following purposes: Verification (to verify a result or an intermediate result during the course of an experiment); Comparison (compare a certain result against an existing result); Progression (analysis of origin of results of an experiment) and Evolution (following the natural course of exploration during an experiment).

The coupling between the results and the associated provenance is inherent thus justifying the development of techniques to facilitate easy viewing of both. Kunde's requirements fail to encompass all aspects of our research primarily due to the fact that these requirements are very generic. Our research includes other requirements such as customized views of scientific data provenance that depend upon user requirement, and/or access privileges.

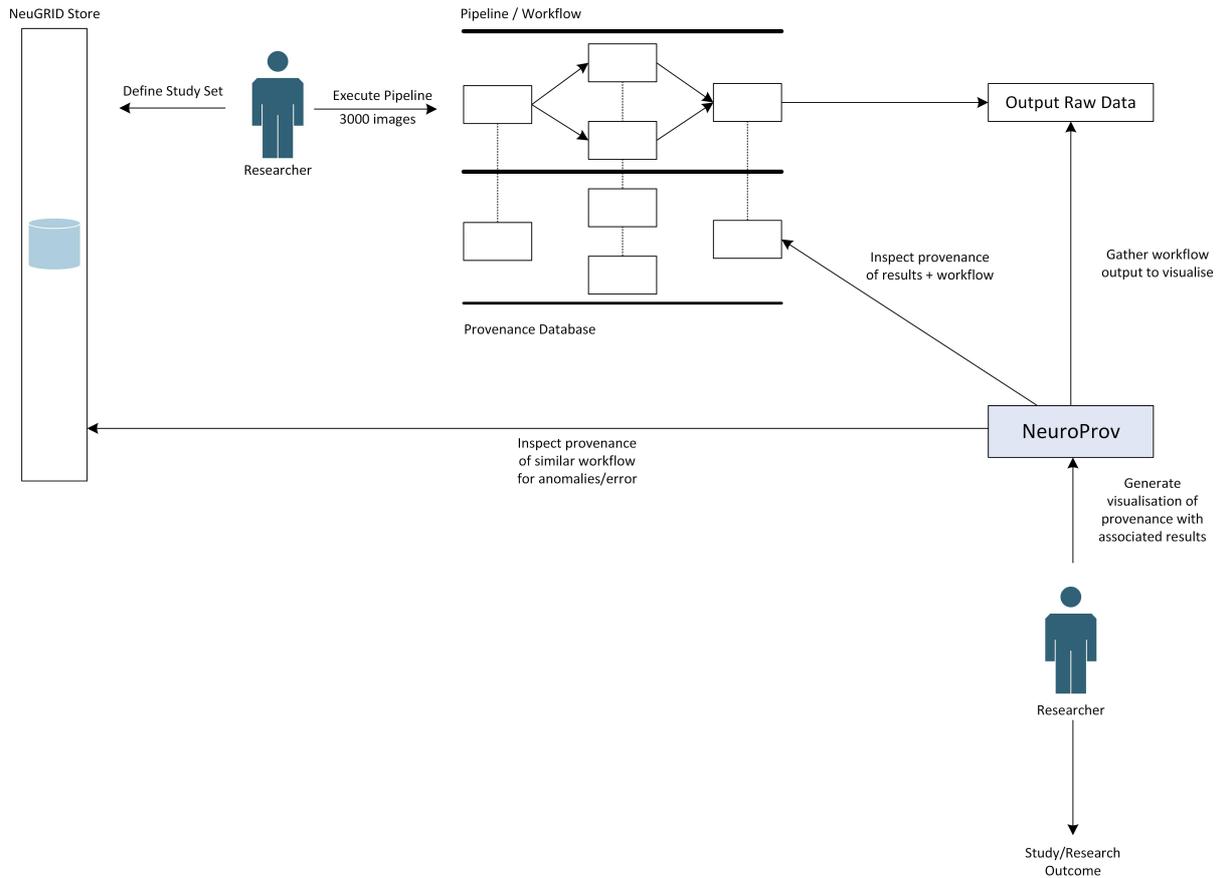

**Figure 1 NeuroProv System Architecture**

## IV NeuroProv System Architecture

This research study is conducted within the context of N4U [9]. N4U or simply neuGRID4U provides neuroscientists and clinician with the ability to perform high-throughput imaging research, and provide neurologists automated diagnostic imaging markers for neurodegenerative diseases such as Alzheimer's for individual patient diagnosis. The experiment also allows users to securely upload, use, share brain scans paired with access to computational power, large image datasets and specialised support and training for conducting neuroimaging analysis. The intended benefit of this project is to enable the discovery of biomarkers for Alzheimer's disease that will improve diagnosis and help speed the development of innovative drugs. Within the context of N4U, end-user community has identified a vital need for provenance. Visualisation of provenance data will allow clinicians and researchers to understand and interpret results from these experiments and provides insight for future research.

Figure 1 presents the system architecture of NeuroProv a visualisation system developed based on user requirements from N4U. The NeuGRID store is a repository of brain scans, associated metadata, workflow execution information, datasets and related provenance information to the above mentioned items. Workflows as defined earlier are a set of tasks in an order to achieve an overall goal. The neuroscientist defines a study set based on the requirement of an analysis selecting the dataset, images to execute. Once the workflow has been executed, data and process provenance are stored in the Provenance database. The researcher can now use NeuroProv to provide a visualisation of provenance based on the workflow executed to verify a result, compare the execution of workflow with a past analysis etc.

This sets the scene for the following section in which the use-cases for NeuroProv are presented to express the need for provenance visualisation for neuroimaging analysis. In N4U, various users such as Research Leaders, Researchers, Pipeline Developers, Image/Data Input Managers and system administrators use provenance data for numerous purposes. For example a workflow yields some surprising and possibly significant results. A research may wish to confirm that the results are accurate and identify any mistakes that may have been made. Visualisation of provenance data for the workflow provides means to analyse all the intermediary image sets and results to verify that the results were incorrect. It may be found that the error was due to a specific group of images interacting badly within the workflow. The user can then annotate the workflow so that other users are warned if they attempt a similar analysis.

Sometimes it may not be enough to reproduce the results. It may also be necessary to validate and, if required, reproduce the workflow that has been used to obtain the results. This makes users confident not only in the results that have been produced but also in the process that led them to generate these results. For example, a user may create a new workflow and run it on a test data set. At each stage in the execution of the workflow, the intermediary images or data are stored and a full provenance track is kept. After results have been produced, the user can examine the visualised provenance to check that each stage of the analysis was completed correctly. The raw results can then be exported into the user's preferred analysis tool and the whole process can be added to the researcher's history for future reference. Initially the new workflow may produce some poor results during testing. The researcher therefore can inspect the visualised provenance of the workflow execution and locate the problem. The user can then interact with the system to make changes to the relevant settings and re-run the test study. This time the process may run correctly and meaningful results may be produced. Without the mechanism to validate workflows, it would not be possible to correct the process and generate accurate results. Therefore visualisation of provenance data helps the researcher to validate results and workflows. The following section illustrates the use-cases defined for NeuroProv and the associated requirements for visualisation of provenance data.

**V NeuroProv Use Cases**

1. Verification: Workflow composition allows users to verify the correctness of a result or an intermediate result.

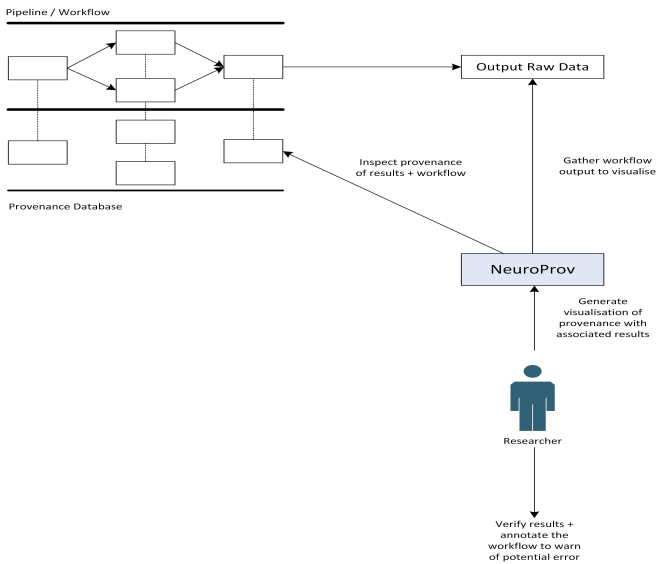

2. Comparison: Insight often comes from comparing different provenance visualisations. Users can perform the experiment with the same attributes using different workflows to compare the results. This will provide further insights in to the experiment.

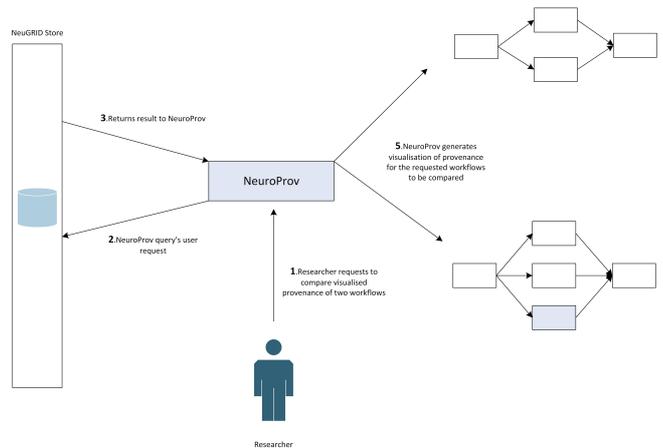

3. Progression: Researchers can view the visualised provenance in order to determine the analysis of origin of results. In a collaborative environment such as NeuGRID scientists frequently work with data that has been collected or processed by other groups or organization. In order to verify the results for correctness, scientists require viewing the progression of the data product with the help of visualisation.

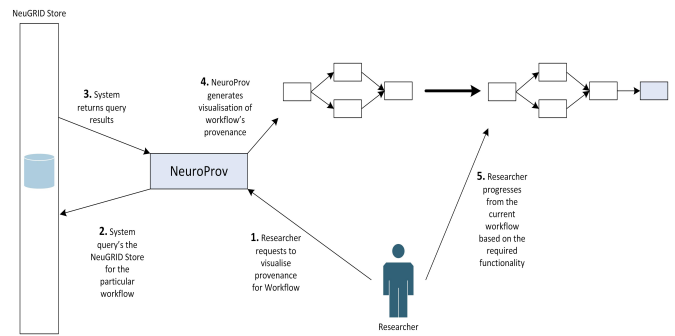

4. Evolution: in order to determine how a certain data product has evolved during the course of the experiment, researchers need to view the visualised provenance.

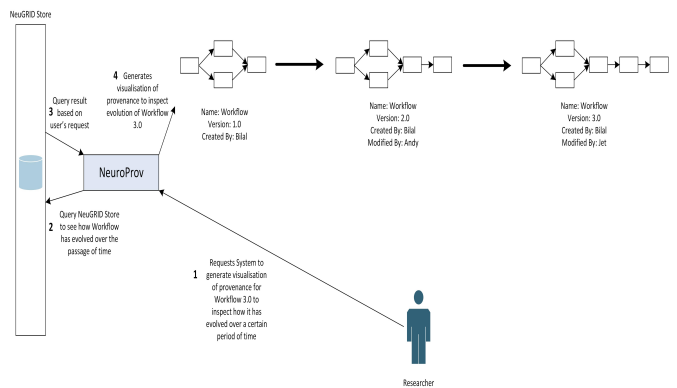

5. Validation of Results: using visualised provenance data to validate any error in intermediate stages of the workflow.

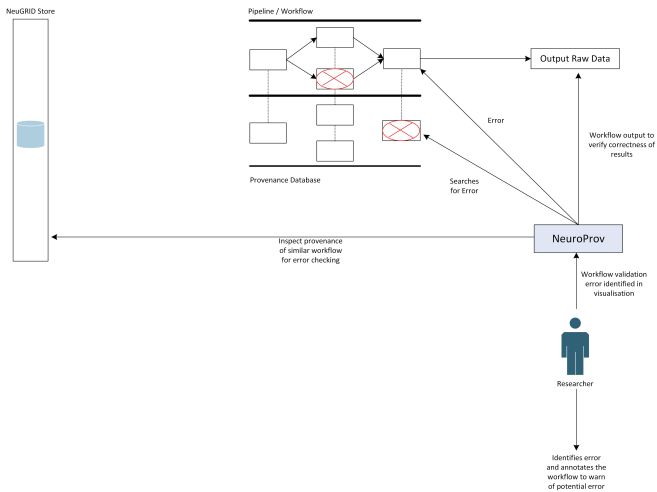

## VI Research Methodology

In order to assist our research, a combination of qualitative and experimental research methodology [21] will be adopted i.e. a subset of the problem is investigated qualitatively through literature review and comparison with existing systems while another subset is analysed empirically through experiments.

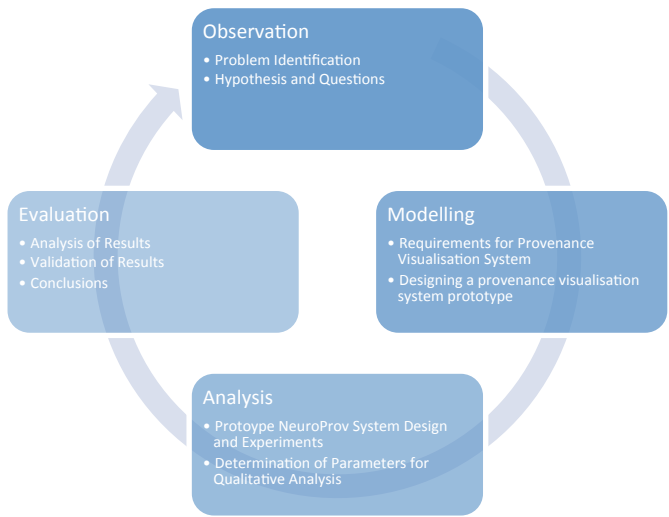

In the Observation phase, the shortcomings of existing provenance visualisation systems are identified from the aspects of collaborative analysis. This step is aided by a thorough literature review leading to the formation of our research goal i.e. visualisation techniques can enhance the utility of provenance data for neuroimaging analysis. Based on the outcome of the observation phase the modelling phase is carried out, which includes determining the requirements of the system that can overcome the limitations identified in the observation phase. The outcome of this phase is a proposed system for visualising provenance data. A prototype system will be developed in the Analysis phase and a set of experiments will be defined, which will be used to test the suitability of the proposed system against the elements of the hypothesis. The criteria for qualitative analysis will also be determined in this phase e.g. characteristics that may define the usability of our provenance visualisation system to a user. The Modelling and Analysis phases will benefit from our interaction with the potential users of the system i.e. scientists from the N4U community.

Results of the analysis phase will be analysed in the evaluation phase and will inform the answers to our research questions and test the validity of our hypothesis. We will also validate the results externally to asses if the results of this research can be generalised for other domains (in addition to neuroimaging related scientific analysis). This reflection on our research results requires iteration of earlier research phases and repetition of experiment steps. Finally, conclusions are drawn from the experimental results and qualitative analysis and the research will conclude by identifying and discussing future directions.

## VII Conclusions and Future Research Directions

The research is under progress and establishes a basis for the need to visualise provenance for neuroimaging analysis. Scientists and researchers need to visualise provenance data in order to aid them in the exploration process by providing means to understand complex data and processes. We have evaluated existing provenance visualisation systems, but through literature survey we have found that they are not adequate for visualising provenance in the domain of neuroimaging analysis. Based on the state-of the-art systems for provenance visualisation we have setup a basis of understanding the need for provenance visualisation for neuroimaging analysis. Provenance visualisation will allow researchers and scientists to work in a collaboratory environment which they can share their findings and mutually benefit from the research. Furthermore it will open future avenues for research for the neuroimaging community. By exploring provenance information using visualisation in a collaboratory manner, scientists can learn by example, expedite their scientific work and potentially reduce time for insight. "The wisdom of the crowd" in context of scientific exploration, can avoid duplication and encourage, documented and reproducible scientific progress.